\documentclass[%
reprint,
aps, pra
]{revtex4-1}

\usepackage{amsmath,amssymb,mathtools}
\usepackage{upgreek}
\usepackage{graphicx}% Include figure files
\usepackage{dcolumn}% Align table columns on decimal point
\usepackage{bm}% bold math
\usepackage{textcomp}
\usepackage{lipsum}
\usepackage{physics}
\usepackage{hyperref}
\usepackage{xcolor}
\hypersetup{colorlinks=true, allcolors=blue}
\usepackage{enumitem}
\usepackage{multirow}

\begin{document}

% on the title page in preprint frequency.
%\preprint{}

\title{Precessional spin-torque dynamics in biaxial antiferromagnets} %Title of paper



\author{Arun Parthasarathy}
\email[]{arun.parth@nyu.edu}
%\thanks{}
\affiliation{\text{Department of Electrical and Computer Engineering, New York University, Brooklyn, New York 11201, USA}}

\author{Egecan Cogulu}
\email[]{egecancogulu@nyu.edu}
%\thanks{}
\affiliation{\text{Center for Quantum Phenomena, Department of Physics, New York University, New York, New York 10003, USA}}

\author{Andrew D. Kent}
\email[]{andy.kent@nyu.edu}
%\thanks{}
\affiliation{\text{Center for Quantum Phenomena, Department of Physics, New York University, New York, New York 10003, USA}}

\author{Shaloo Rakheja}
\email[]{rakheja@illinois.edu}
%\thanks{}
\affiliation{Holonyak Micro and Nanotechnology Laboratory, University of Illinois at Urbana-Champaign, Urbana, IL 61801}


\date{\today}

\begin{abstract}
The N\'eel order of an antiferromagnet subject to a spin torque can undergo precession in a circular orbit about any chosen axis. To orient and stabilize the motion against the effects of magnetic anisotropy, the spin polarization should have components in-plane and normal to the plane of the orbit, where the latter must exceed a threshold. For biaxial antiferromagnets, the precessional motion is described by the equation for a damped-driven pendulum, which has hysteresis a function of the spin current with a critical value where the period diverges. The fundamental frequency of the motion varies inversely with the damping, and as $(x^p-1)^{1/p}$ with the drive-to-criticality ratio $x$ and the parameter $p>2$. An approximate closed-form result for the threshold spin current is presented, which depends on the minimum cutoff frequency the orbit can support.  Precession about the hard axis has zero cutoff frequency and the lowest threshold, while the easy axis has the highest cutoff. A device setup is proposed for electrical control and detection of the dynamics, which is promising to demonstrate a tunable terahertz nano-oscillator.
\end{abstract}

\maketitle %\maketitle must follow title, authors, abstract


\section{Introduction}

In antiferromagnetic materials, injection of a spin current exerts spin torque on alternating spin moments, which can excite steady-state precession of the N\'eel order at terahertz frequencies~\cite{gomonay2014spintronics}. The precession can reciprocally pump spins into an adjacent conductor~\cite{vaidya2020subterahertz}, which may generate an oscillating sinusoidal~\cite{khymyn2017antiferromagnetic} or spike-like~\cite{khymyn2018ultra} electric signal. These electrically-induced antiferromagnetic oscillations could serve as compact narrowband terahertz sources and spike generators for applications in terahertz imaging and sensing~\cite{mittleman2017perspective}, and in neuromorphic computing~\cite{torrejon2017neuromorphic}.

Given a source of spin current, how does the spin torque affect the dynamics of the N\'eel-order? In biaxial-anisotropy antiferromagnets, this problem has been studied for special cases where the spin current is polarized either along the easy or hard anisotropy axis~\footnote{Algebraic expression for threshold drive is known in the limit of small~\cite{khymyn2017antiferromagnetic} and large~\cite{khymyn2018ultra} damping. An expression for the time dependence of the angular speed was found for large drive and small damping~\cite{khymyn2017antiferromagnetic}. The influence of dissipative spin-orbit torque \cite{troncoso2019antiferromagnetic} was studied easy-plane anisotropy.}. The description of the motion for a general angle of spin polarization and parametric study of the nonlinear dynamics are, however, unexplored, which we we analyze in this paper in detail. This topic has
implications for a tunable terahertz detector~\cite{gomonay2018narrow}, which can filter frequencies higher than a certain cutoff frequency determined by the precession axis.

We study the precession of N\'eel order caused by spin torque about an arbitrary rotational axis. The equation of motion (Sec.~\ref{Sec:theory}) permits steady-state solutions where the N\'eel order precesseses in a circular orbit (Sec.~\ref{Sec:analysis}). The dynamics is analyzed separately into angular motion in the orbit which determines the frequency~(Sec.~\ref{Sec:angular}), and perturbation from the orbital plane which determines the stability~(Sec:~\ref{Sec:stability}).  A schema for electrical control and detection of antiferromagnetic dynamics is presented in Sec.~\ref{Sec:electrical}.

\section{Theoretical framework}
\label{Sec:theory}
Antiferromagnets with a bipartite collinear ordering~\cite{dombre1989nonlinear} are composed of two interpenetrating square lattices $A$ and $B$ with oppositely aligned moments. For such ordering, the thermodynamic state of the antiferromagnet in a mean-field approximation is represented by the unit vectors $\vb{m}_{A}$ and $\vb{m}_{B}$ for the sublattice spin moments, each of which has a magnetization $M_\text{s}$~\cite{kittel2004introduction}. There are three significant contributions to the free energy density of an antiferromagnet: the exchange coupling of neighboring spins, the magnetocrystalline anisotropy, and the Zeeman interaction of magnetic moments with an external magnetic field.

\begin{figure}
 \centering
 \includegraphics{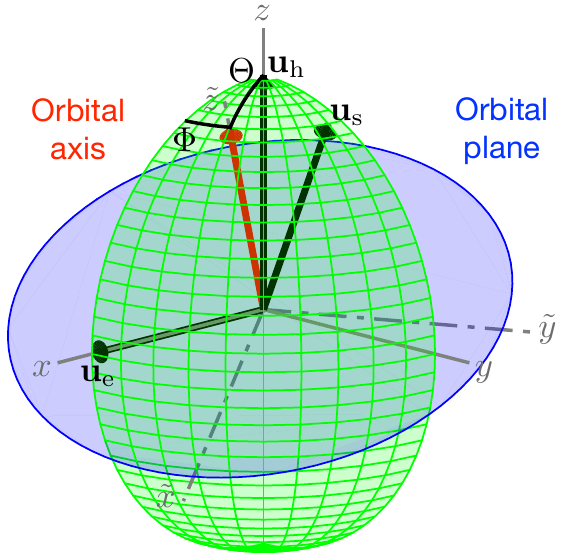}%
 \caption{Representation of magnetic-anisotropy easy axis $\vb{u}_\text{e}$ and hard axis $\vb{u}_\text{h}$, spin-polarization direction $\vb{u}_\text{s}$, and steady-state circular orbit of N\'eel order on a unit-sphere quadrant.}%
 \label{fig:UnitVec}
 \end{figure}

Biaxial-anisotropy antiferromagnets are characterized by a direction of minimum energy which is `easy' for spins to orient along, and an orthogonal direction of maximum energy which is `hard'. Consider a monodomain film of thickness $d_\text{a}$ with the easy and hard axes oriented along the unit vectors $\vb{u}_\text{e}$ and $\vb{u}_\text{h}$, respectively (Fig.~\ref{fig:UnitVec}). If the antiferromagnetic exchange coupling between the sublattice moments is much stronger than the anisotropy, the average net moment $\vb{m} = (\vb{m}_{A} + \vb{m}_{B})/2$ and the N\'eel order $\vb{n}=(\vb{m}_{A} - \vb{m}_{B})/2$ satisfy the relation $\vb{n}^2 = 1 - \vb{m}^2  \approx 1$. The total free energy density in the presence of an external magnetic field $H$ applied along the unit vector $\vb{h}$ up to lowest order is~\cite{qaiumzadeh2017spin, parthasarathy2019phenomenological}
\begin{equation}
\mathcal{F} = \mathcal{J}\vb{m}^2 + \mathcal{K}_\text{h} (\vb{n}\cdot \vb{u}_\text{h})^2 -\mathcal{K}_\text{e} (\vb{n}\cdot \vb{u}_\text{e})^2 - 2\mathcal{Z} \vb{m}\cdot\vb{h},
\label{eq:Fmn}
\end{equation}
where the exchange coupling $\mathcal{J}$, the anisotropy coefficients $\mathcal{K}_i$ and the Zeeman energy density $\mathcal{Z}=\mu_0M_\text{s}H$ are positive numbers, and $\mathcal{J}\gg\mathcal{K}_i, \mathcal{Z}$.

In the limit of large exchange, the problem of coupled dynamics of the sublattice moments approximately reduces to motion of the N\'eel order on a unit-sphere phase space~\cite{KOSEVICH1990117, Gomonay2010}. Starting from the Landau-Lifshitz-Gilbert equation augmented with the antidamping-like torque~\cite{SLONCZEWSKI1996L1, sinova2015, jungwirth2016antiferromagnetic} induced by a spin current, the $\sigma$-model equation of motion for the N\'eel order is~\cite{KOSEVICH1990117, parthasarathy2019phenomenological}
\begin{align}
&\vb{n}\times\bigg[\frac{M_\text{s}}{\gamma\mathcal{J}}\ddot{\vb{n}} + \alpha\dot{\vb{n}} + \frac{\gamma}{M_\text{s}}(\mathcal{K}_\text{h} (\vb{n}\cdot\vb{u}_\text{h}) \vb{u}_\text{h} - \mathcal{K}_\text{e} (\vb{n}\cdot\vb{u}_\text{e}) \vb{u}_\text{e}) +  \nonumber \\ 
&\frac{\mathcal{Z}}{\mathcal{J}}\big(2\dot{\vb{n}}\times\vb{h} + \gamma\mu_0 H(\vb{n}\cdot\vb{h})\vb{h} \big)
+ \frac{\gamma J_\text{s}}{M_\text{s} d_\text{a}} (\vb{n}\times\vb{u}_\text{s}) \bigg] = 0,
\label{eq:eom}
\end{align}
where an overdot denotes derivative with respect to time $t$, $\alpha>0$ is the Gilbert damping, $\gamma$ is the absolute gyromagnetic ratio, and $J_\text{s}$ is the injected spin-current density with non-equilibrium spin polarization along the unit vector $\vb{u}_\text{s}$~\footnote{$J_\text{s}>0$ in the sense of electrons with spin-polarization direction $\vb{u}_\text{s}$ (or $-\vb{u}_\text{s}$) injecting into (or ejecting from) the magnetic material.}. The equation is second order, implying that N\'eel-order has inertia. The net moment~\cite{parthasarathy2019phenomenological}
\begin{equation}
\vb{m} = \frac{M_\text{s}}{\gamma\mathcal{J}}(\dot{\vb{n}}\times\vb{n}) + \frac{\mathcal{\mathcal{\mathcal{Z}}}}{\mathcal{J}}[\vb{h}-(\vb{n}\cdot\vb{h})\vb{n}]
\end{equation}
becomes a dependent variable, allowing for the N\'eel order dynamics \eqref{eq:eom} to be solved independently.

Multiplying Eq.~\eqref{eq:eom} by $M_\text{s}/(\gamma \mathcal{K}_\text{e})$, when the external field is much smaller than spin-flop field $\mathcal{\mathcal{Z}} \ll \sqrt{\mathcal{J}\mathcal{K}_\text{e}}$~\footnote{The spin-flop field is around $\sim10$ T for antiferromagnetic insulators such as NiO~\cite{wiegelmann1994magnetoelectric}, Cr\textsubscript{2}O\textsubscript{3}~\cite{machado2017spin} and MnF\textsubscript{2}~\cite{jacobs1961spin}.}, the $\mathcal{Z}$-coefficient terms need not be considered. In dimensionless form, where the time is scaled by characteristic frequency $\tau_0 = t(\gamma\sqrt{\mathcal{J}\mathcal{K}_\text{e}}/M_\text{s})$, the equation of motion becomes
\begin{align}
\vb{n}\times[\vb{n}'' + \beta_0\vb{n}' + \kappa(\vb{n}\cdot\vb{u}_\text{h}) \vb{u}_\text{h} &- (\vb{n}\cdot\vb{u}_\text{e}) \vb{u}_\text{e}   \nonumber \\ 
&  + \Gamma_0(\vb{n}\times\vb{u}_\text{s})/2] = 0,
\label{eq:eqomnd}
\end{align}
where a prime denotes derivative with respect to $\tau_0$, and the dimensionless parameters are defined as $\beta_0 = \alpha\sqrt{\mathcal{J}/\mathcal{K}_\text{e}}$, $\kappa = \mathcal{K}_\text{h}/\mathcal{K}_\text{e}$ and $\Gamma_0 = 2J_\text{s}/(\mathcal{K}_\text{e} d_\text{a})$. From hereafter, $\Gamma_0\vb{u}_\text{s}$ is called the spin-polarization vector.

\section{Analysis of solution} 
\label{Sec:analysis}

In steady state, Eq.~\eqref{eq:eqomnd} can have stationary solutions $\vb{n}'=0$ and time-varying solutions $\vb{n}'\neq 0$ depending on the parameters and initial condition,. Vector product of Eq.~\eqref{eq:eqomnd} with the angular speed $\vb{n}\times\vb{n}'$ followed by dot product with $\vb{n}$ produces
\begin{align}
\kappa(\vb{n}\cdot\vb{u}_\text{h})[(&\vb{n}\times\vb{n}')\cdot\vb{u}_\text{h}] 
- (\vb{n}\cdot\vb{u}_\text{e})[(\vb{n}\times\vb{n}')\cdot\vb{u}_\text{e}] \nonumber \\ &+ (\vb{n}\times\vb{n}')\cdot\vb{n}'' + \Gamma_0(\vb{n}'\cdot\vb{u}_\text{s})/2 = 0,
\label{eq:relation}
\end{align}
 the relationship between the projection of $\vb{u}_\text{e}$, $\vb{u}_\text{h}$ and $\vb{u}_\text{s}$ on the orthogonal vectors $\vb{n}$, $\vb{n}'$ and $\vb{n}\times\vb{n}'$.
 
Among time-varying solutions, we claim that $(\vb{n}\times\vb{n}')\cdot\vb{n}'' =0$ is a subset wherein the N\'eel order precesses in a circular orbit around the origin. Let the orbital axis make a polar angle $\Theta$ from the hard axis along $z$ and an azimuthal angle $\Phi$ from the easy axis along $x$ as shown in Fig.~\ref{fig:UnitVec}, with intermediate axis along $y$. The orthonormal basis suitable for the orbital plane is defined as $\vb{\tilde{x}} = [\cos\Theta\cos\Phi, \cos\Theta\sin\Phi, -\sin\Theta]$, $\vb{\tilde{y}} = [-\sin\Phi, \cos\Phi, 0]$ and $\vb{\tilde{z}} = [\sin\Theta\cos\Phi, \sin\Theta\sin\Phi, \cos\Theta]$. If the N\'eel order makes an angle $\phi$ counterclockwise from $\vb{\tilde{x}}$ axis in the orbital plane and rotates with angular speed $\phi'$, $\vb{n} = \vb{\tilde{x}}\cos\phi + \vb{\tilde{y}}\sin\phi$ and $\vb{n}' = \phi'(-\vb{\tilde{x}} \sin\phi  + \vb{\tilde{y}}\cos\phi) $. On substitution into Eq.~\eqref{eq:relation}, the orbital axis orients along $(\Theta, \Phi)$ if the components of the spin-polarization vector in the orbital plane satisfy
\begin{subequations}
\begin{align}
\Gamma_0(\vb{\tilde{x}}\cdot \vb{u}_\text{s}) &= \sin(\Theta)\sin(2\Phi),\\
\Gamma_0(\vb{\tilde{y}}\cdot \vb{u}_\text{s}) &= (\kappa+\cos^2\Phi)\sin(2\Theta),
\end{align}
\label{eq:oao}%
\end{subequations}
and $\phi'\neq 0$, which requires the perpendicular component $\Gamma_0(\vb{\tilde{z}}\cdot \vb{u}_\text{s})$ to overcome a certain threshold to drive the motion against damping. Determining this condition requires analyzing the angular motion and stability of the orbiting N\'eel order.

\subsection{Angular motion in orbit}
\label{Sec:angular}

Equation~\eqref{eq:eqomnd} implies that tangential components of $\vb{n}$ inside the square bracket should vanish. Equating the component $\boldsymbol{\phi} = -\vb{\tilde{x}} \sin\phi  + \vb{\tilde{y}}\cos\phi$ to zero gives the angular motion in the orbital plane 
\begin{align}
2\phi'' &+ 2\beta_0\phi' + [\cos(2\Phi) - (\kappa+\cos^2\Phi)\sin^2\Theta]\sin(2\phi) \nonumber \\
&+ [\cos\Theta\sin(2\Phi)]\cos(2\phi) = \Gamma_0(\vb{\tilde{z}}\cdot \vb{u}_\text{s}).
\end{align}
Combining $\sin(2\phi)$ and $\cos(2\phi)$ terms together, and performing the following replacements 
\begin{gather}
C = \cos(2\Phi) - (\kappa+\cos^2\Phi)\sin^2\Theta,\quad D = \cos\Theta\sin(2\Phi); \nonumber  \\
g= \sqrt{C^2+D^2},\quad \beta = \frac{\beta_0}{\sqrt{g}},\quad \tau = \frac{\tau_0}{\sqrt{g}}\quad \Gamma = \frac{\Gamma_0\abs{\vb{\tilde{z}}\cdot \vb{u}_\text{s}}}{g}; \nonumber  \\
\varphi=\mathrm{sgn}(\vb{\tilde{z}}\cdot \vb{u}_\text{s})\left[2\phi + \arctan(\frac{D}{C})\right];
\label{eq:repl}
\end{gather}
recast the equation of motion into (derivative denoted by prime for $\varphi$ are with respect to $\tau$)
\begin{equation}
\varphi'' + {\beta}\varphi' + \sin\varphi = {\Gamma},
\label{eq:ddpend}
\end{equation}
that of a unit-mass pendulum of unit length in effective gravity $g$ with a viscous damping ${\beta}$ and driven by a constant tangential force ${\Gamma}$~\cite{coullet2005damped}.

\begin{figure}
 \centering
 \includegraphics{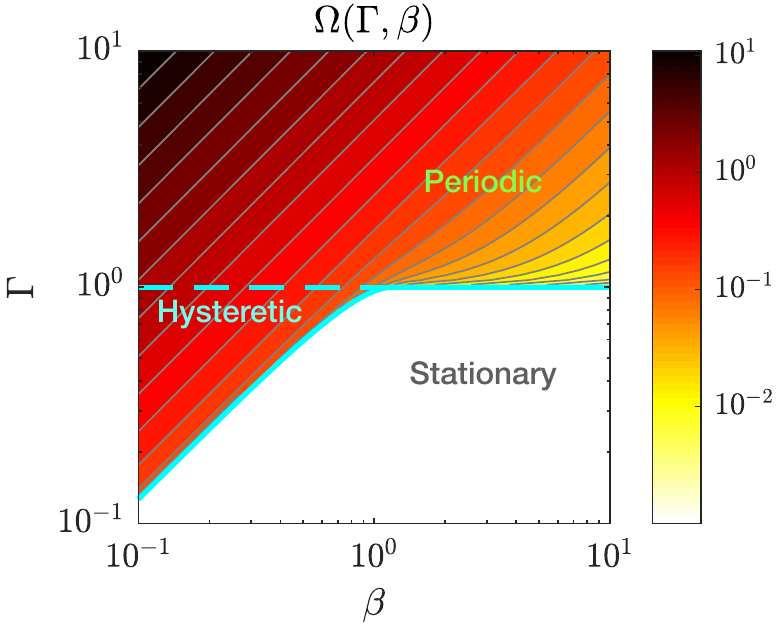}%
 \caption{Fundamental frequency $\Omega$~\eqref{eq:Omega_general} of the damped-driven pendulum equation \eqref{eq:ddpend} in the parameter space of effective damping $\beta$ and drive $\Gamma$ represented by contour map with values on colorbar.}%
 \label{fig:conplot}
 \end{figure}

The solution of the damped-driven pendulum equation~\eqref{eq:ddpend} depends only on two dimensionless parameters $\beta$ and $\Gamma$. The steady-state dynamics of this non-linear system yields either a stationary solution [$\varphi'(\tau)=0$], or a periodic solution [$\varphi'(\tau+T) = \varphi'(\tau)\neq 0$] with a period $T$.  Transient simulation with initial conditions $\varphi_0 = \pi$ and $\varphi'_0 = 0$ over the parameter space is used to obtain the time $T$ to complete a revolution in steady state ($\tau > 7/\beta$). A contour plot of the fundamental frequency is shown in Fig.~\ref{fig:conplot} depicting the regions of different types of solutions. 

In the hysteretic region, the steady-state solution can be periodic or stationary depending on the initial condition. For example, if the pendulum was released vertically upside down ($\varphi_0=\pi$), then it revolves continually; but if it was released horizontally from a point where the gravity opposes the drive ($\varphi_0=\pi/2$), then it dips and settles to equilibrium at $\sin\varphi_\text{eq} = \Gamma$. The hysteretic region is defined as: $f_\text{h}(\beta)< \Gamma(\beta) < 1$ on the domain $0<\beta<\beta_\text{h}$, where $f_\text{h}$ determines the homoclinic bifurcation~\cite{strogatz2018nonlinear}. The data is fit by a cubic-polynomial
\begin{equation}
f_\text{h}(\beta) = a_0\beta+ \left(\frac{3-2 a_0 \beta_\text{h}}{\beta_\text{h}^2}\right)\beta^2 + \left(\frac{a_0 \beta_\text{h} - 2}{\beta_\text{h}^3}\right)\beta^3,
\end{equation}
where $a_0 = 4/\pi$~\cite{coullet2005damped} and $\beta_\text{h}\approx 1.2$. The minimum threshold drive needed to sustain revolutions (solid cyan line) is expressed 
\begin{equation}
\Gamma_\text{m}(\beta) = 
\begin{cases} 
      f_\text{h}(\beta) & \beta < \beta_\text{h} \\
      1 & \beta \geq \beta_\text{h}
\end{cases},
\label{eq:Gammam}
\end{equation}
where $\Gamma_\text{m}=1$ is the threshold that can initiate revolutions independent of the damping (dashed cyan line).

In the overdrive limit $\Gamma\gg \Gamma_\text{m} $, the fundamental frequency is calculated by integrating Eq.~\eqref{eq:ddpend} over a period and ignoring the role of gravity ($\sin\varphi$ term) to obtain
\begin{equation}
\Omega_{\text{O}\Gamma} =  \frac{1}{2\pi} \frac{\Gamma}{\beta}.
\label{eq:fOD}
\end{equation}
This is consistent with the slope of the contour lines seen in Fig.~\ref{fig:conplot} in regions far above the threshold. The time-dependence of the steady-state angular speed $\omega=\varphi'$ is estimated from Eq.~\eqref{eq:ddpend} by truncating $\sin\varphi \simeq \sin(\Gamma \tau\beta)$, multiplying by $e^{\beta\tau}$, integrating and letting the transients decay to give
\begin{equation}
\omega_{\text{O}\Gamma} (\tau) \simeq \frac{\Gamma}{\beta} - \frac{\beta}{\sqrt{\beta^4+\Gamma^2}}\sin(\frac{\Gamma}{\beta}\tau - \xi_{\text{O}\Gamma}),
\label{eq:omegaOD}
\end{equation}
where $\xi_{\text{O}\Gamma} = \arctan(\Gamma/\beta^2)$.

In the overdamped limit $\beta\gg1$, Eq.~\eqref{eq:ddpend} reduces to $ \varphi' =  (\Gamma-\sin\varphi)/\beta$, which is separable and has a closed-form integral. The periodic solution is expressed~\cite{coullet2005damped}
\begin{equation}
\Gamma\tan\frac{\varphi_{\text{O}\beta}(\tau)}{2} =1 + \sqrt{\Gamma^2-1}\tan(\frac{\sqrt{\Gamma^2-1}}{2\beta}\tau - \xi_{\text{O}\beta}), \nonumber
\end{equation}
where $\xi_{\text{O}\beta} = \arctan(1/\sqrt{\Gamma^2-1})$. The fundamental frequency is  
\begin{equation}
\Omega_{\text{O}\beta} = \frac{1}{2\pi}\frac{\sqrt{\Gamma^2-1}}{\beta},
\label{eq:Omega_od}
\end{equation}
and the steady-state angular speed is  
\begin{equation}
\omega_{\text{O}\beta}(\tau) = \frac{\Gamma}{\beta} - \frac{2}{\beta}\frac{\tan(\varphi_{\text{O}\beta}/2)}{1+\tan^2(\varphi_{\text{O}\beta}/2)}.
\label{eq:omegaod}
\end{equation}
Superposing the overdrive limit on Eq.~\eqref{eq:omegaod} results in convergence of solutions with Eq.~\eqref{eq:omegaOD}.

\begin{figure}
\centering
\includegraphics{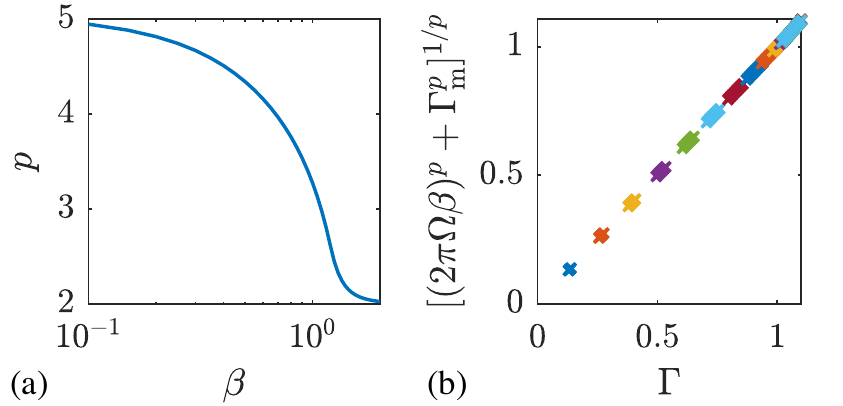}%
\caption{Plot of Eq.~\eqref{eq:Omega_general}. (a) The parameter $p$ increases from 2 for $\beta \lesssim 1$. (b) The data for $\Omega$ evaluated with the vertical-axis function fits accurately with $\Gamma$, where adjoining markers of the same color have a unique $\beta$ (decreasing diagonally up).}%
\label{fig:freqscaling}
\end{figure}

For moderate drive $\Gamma\gtrsim\Gamma_\text{m}$ and damping $\beta \lesssim 1$, it is not possible to derive analytic solutions~\cite{coullet2005damped, gitterman2008noisy}. However, the fundamental frequency closely follows the law
\begin{equation}
\Omega(\Gamma,\beta) = \frac{1}{2\pi}\frac{\big[\Gamma^{p} - \Gamma_\text{m}^{p}\big]^{1/p}}{\beta} \quad\quad 
\Bigg\{\begin{matrix*}[l] 
\Gamma_\text{m}= \Gamma_\text{m}(\beta) \\[1pt] 
p = p(\beta)
\end{matrix*},
\label{eq:Omega_general}
\end{equation}
conceived by generalizing the overdamped result~\eqref{eq:Omega_od}. The functional relationship $p(\beta)$ and the accuracy of the model are depicted in Fig.~\ref{fig:freqscaling}. The inverse dependence of $p$ on $\beta$ implies that a weakly damped system $\beta\ll 1$ is much easier to oscillate than a strongly damped $\beta\gg 1$ for a fixed drive-to-threshold ratio. Superposing the overdrive limit results in convergence with Eq.~\eqref{eq:fOD}.

\begin{figure}
\centering
\includegraphics{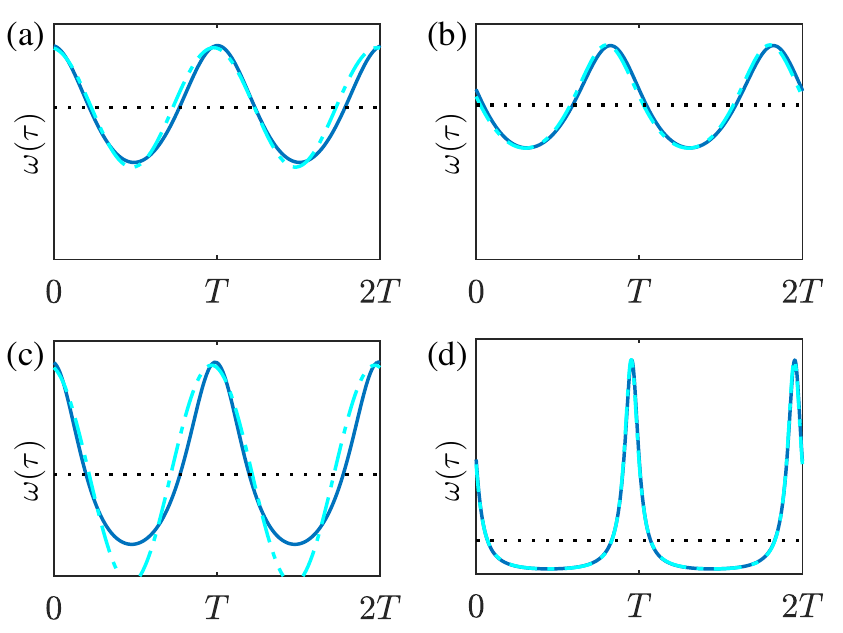}%
\caption{Waveforms of the steady-state angular speed $\omega(\tau)$ of the damped-driven pendulum revolving with time period $T$ for (a) overdrive $\epsilon_{\text{O}\Gamma}=0.05$ and small damping $\beta=0.2$, (b) overdrive $\epsilon_{\text{O}\Gamma}=0.05$ and large damping $\beta=5$, (c) near-threshold drive $\epsilon_{\text{th}} = 0.05$ and small damping $\beta=0.2$, (d) near-threshold drive $\epsilon_{\text{th}} = 0.05$ and large damping $\beta=5$. Dark-color plot is the exact numerical solution and light-color plot is the approximate closed-form solution. Dotted horizontal line denotes the average value equal to $2\pi\Omega$~\eqref{eq:Omega_general}. At time $\tau=0$, the pendulum is at the lowest point $\varphi=0$.}%
\label{fig:transwaveforms}
\end{figure}

To compare periodic solutions across the parameter space, the first-order term $\epsilon_{\text{O}\Gamma} = (\Gamma_\text{m}/\Gamma)^{p}/p$ in the expansion of $2\pi\beta\Omega/\Gamma$ \eqref{eq:Omega_general} is used to measure nearness to overdrive; the relative increment $\epsilon_{\text{th}} = (\Gamma-\Gamma_\text{m})/\Gamma_\text{m}$ is used to measure nearness to threshold. The waveforms of the angular speed for limiting cases of damping and drive are juxtaposed in Fig.~\ref{fig:transwaveforms}. The closed-form solutions agree with the numerical ones, except when the drive is near the minimum threshold and the damping is weak. The oscillations in angular speed are coherent in the overdrive limit, and Dirac-comb-like in the overdamped limit for near-threshold drive.

\subsection{Stability of orbit}
\label{Sec:stability}

Orbits that are elevated from the plane perpendicular to hard axis may become stable only above a critical angular speed~\footnote{Similar to the Round Up amusement ride which has to spin fast enough when the circular horizontal platform rises, so that the centrifugal force is able to push riders against the wall and prevent them from falling.}. To study stability, the N\'eel order is perturbed from the orbital plane by a small angle $\delta\theta$ or $\vb{n} = \vb{\tilde{x}}\cos\phi + \vb{\tilde{y}}\sin\phi + \vb{\tilde{z}}\delta\theta$.  Equating the component $\boldsymbol{{\theta}}=(\vb{\tilde{x}}\cos\phi + \vb{\tilde{y}}\sin\phi)\delta\theta - \vb{\tilde{z}}$ inside the square bracket of Eq.~\eqref{eq:eqomnd} to zero, using the relation of Eq.~\eqref{eq:oao}, and retaining only linear terms in $\delta\theta$ gives the equation of a damped harmonic oscillator $\delta \theta'' + \beta_0 \delta\theta' + K \delta\theta = 0$, where the spring constant
\begin{align}
K &= \kappa(\cos^2\Theta - \sin^2\Theta\cos^2\phi) - \sin^2\Theta\cos^2\Phi  \nonumber \\
&+ (\cos\Theta\cos\Phi\cos\phi-\sin\Phi\sin\phi)^2 +\phi'^2. 
\label{eq:K}
\end{align}

\begin{figure*}
\centering
\includegraphics{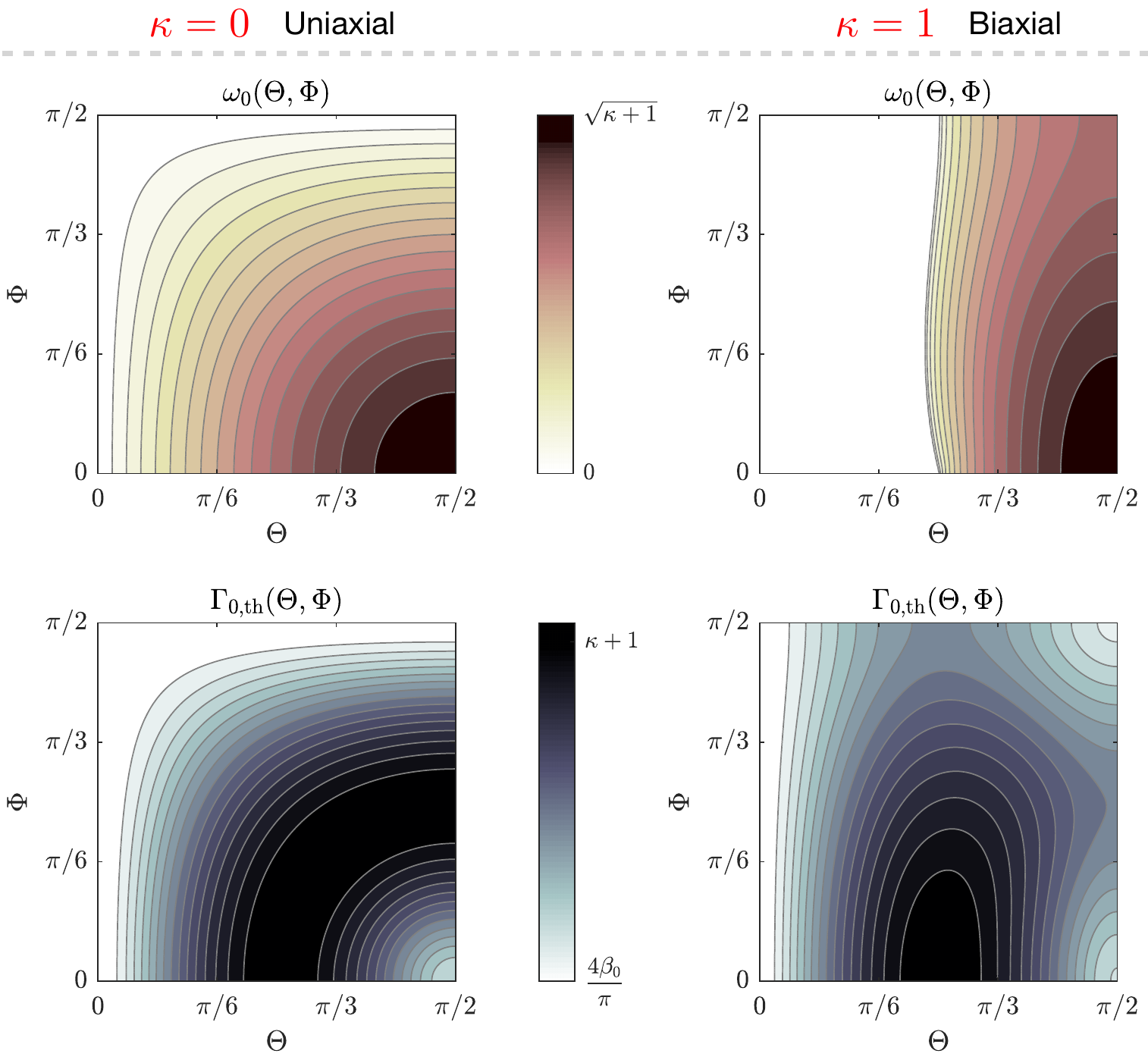}%
\caption{The minimum angular frequency $\omega_0$ (top) and threshold spin current $\Gamma_{0,\text{th}}$ (bottom) which can orient and stabilize the orbital motion of the N\'eel order about an axis $(\Theta, \Phi)$ when damping is small $\beta_0=0.2$ for uniaxial (left) and biaxial (right) anisotropy, represented by contour map with range of values on the colorbar (middle). The minimum frequency $\omega_0$ increases with deviation from the hard axis $(\Theta,\Phi) = (0,\Phi)$ and peaks along the easy axis $(\Theta,\Phi) = (\pi/2,0)$. The threshold is lowest for motion about the hard axis, and peaks for an intermediate orientation in-between hard and easy axis.}%
\label{fig:Gamma0th}
\end{figure*}

To understand how $K$ depends on orbital parameters~\eqref{eq:repl}, expression~\eqref{eq:K} is expanded and compactly rewritten
\begin{equation}
K = \kappa-1-3\mathcal{E} + \frac{g}{2}\left(\cos\varphi + \frac{\varphi'^2}{2}\right),
\end{equation}
where the angle-averaged energy of the orbit
\begin{align}
\mathcal{E} &= \ev{\kappa(\vb{n}\cdot \vb{u}_\text{h})^2 - (\vb{n}\cdot \vb{u}_\text{e})^2} \nonumber \\
&= \frac{1}{2}[(\kappa+\cos^2\Phi)\sin^2\Theta - 1].
\end{align}
If $K$ is positive throughout the angular motion, the N\'eel order will restore to the orbital plane. The precise condition for stability for a general $\beta$ requires numerically evaluating the steady-state value of $(\cos\varphi + \varphi'^2/2)$ in the parameter space of $\beta$ and $\Gamma$ (similar to how fundamental frequency was obtained in Sec.~\ref{Sec:angular}), and checking if $\min_\varphi K > 0$. But, an approximate closed-form result is obtained by treating $\varphi'$ as a constant average value $2\pi\Omega$~\eqref{eq:Omega_general}, which is accurate in the overdrive limit. Then, the stability condition is $\min_\varphi K = g\pi^2\Omega^2 - (g/2+3\mathcal{E}+1-\kappa) > 0$, which gives the minimum angular frequency
\begin{equation}
\omega_0 = \Re\big(\sqrt{g/2+3\mathcal{E}+1-\kappa}\ \big),
\end{equation}
which is positive for orientations satisfying $g/2+3\mathcal{E}+1-\kappa >0$ (Fig.~\ref{fig:Gamma0th}).
The stability threshold for perpendicular projection of the spin-polarization vector is
\begin{equation}
(\Gamma_0\abs{\vb{\tilde{z}}\cdot \vb{u}_\text{s}})_\text{th} = \left[\left(2\beta_0 \omega_0 \right)^p + \left\{g\Gamma_\text{m}(\beta_0/\sqrt{g})\right\}^p \right]^{1/p},
\end{equation}
where $p = p(\beta_0/\sqrt{g})$ (Fig.~\ref{fig:freqscaling}a).

In essence, to orient and stabilize the orbital motion about an axis $(\Theta, \Phi)$, the spin-polarization vector 
\begin{align}
\Gamma_{0}\vb{u}_\text{s} = \sin(\Theta)\sin(2\Phi) \vb{\tilde{x}} &+ (\kappa +\cos^2\Phi)\sin(2\Theta) \vb{\tilde{y}} \nonumber \\
&+\Gamma_0(\vb{\tilde{z}}\cdot \vb{u}_\text{s}) \vb{\tilde{z}},
\label{eq:Gamma0vec} 
\end{align}
where $\Gamma_0$ should exceed the threshold
\begin{align}
\Gamma_{0,\text{th}} = \big[\sin^2(\Theta)\sin^2(2\Phi) &+ (\kappa +\cos^2\Phi)^2\sin^2(2\Theta)  \nonumber \\
& + (\Gamma_0\abs{\vb{\tilde{z}}\cdot \vb{u}_\text{s})})_\text{th}^2 \big]^{1/2}.
\label{eq:Gamma0th}
\end{align}
The spin-polarization direction does not coincide with the orbital axis, except for directions along the anisotropy axes for which $\Gamma_{0,\text{th}}=(\Gamma_0\abs{\vb{\tilde{z}}\cdot \vb{u}_\text{s}})_\text{th}$. The orbital parameters to orient along the anisotropy axes are as follows: hard axis has $(\Theta,\Phi) = (0,\Phi)$, $g=1$ and $\omega_0=0$, intermediate axis has $(\Theta,\Phi) = (\pi/2,\pi/2)$, $g=\kappa+1$ and $\omega_0=\sqrt{\kappa}$, and easy axis has $(\Theta,\Phi) = (\pi/2,0)$, $g=\kappa$ and $\omega_0=\sqrt{\kappa+1}$. The hard axis supports orbit with the lowest threshold and minimum angular frequency, while the easy axis has the highest cutoff. The axis along $(\Theta,\Phi) = (\arctan(1/\sqrt{\kappa}),0)$ has $g=0$ meaning no oscillations in angular speed. For small damping, the range of threshold varies from $4\beta_0/\pi$ to $\kappa+1$ (Fig.~\ref{fig:Gamma0th}).  

Below the threshold spin current, for moderate values of damping, the N\'eel-order motion can evolve into chaos (aperiodic long-term behavior) for a small window of current, before settling down to equilibrium, which will be the subject of future research.

Thus, the choice of the orbital axis, and the parameters of damping and anisotropy determine the threshold current, cutoff frequency and waveform of the angular speed of N\'eel order. These mathematical results may be translated into electrically measurable input and output signals via the device setup presented in the next section.

\section{Electrical control and detection}
\label{Sec:electrical}

Electrical means of controlling the spin current injected into the antiferromagnet and detecting the resulting oscillations in a thin-film system are fundamental to spintronic devices and applications in technology. In this regard, the setup should be all-electric with controllable spin-polarization direction.

\begin{figure}
\centering
\includegraphics[width=\linewidth]{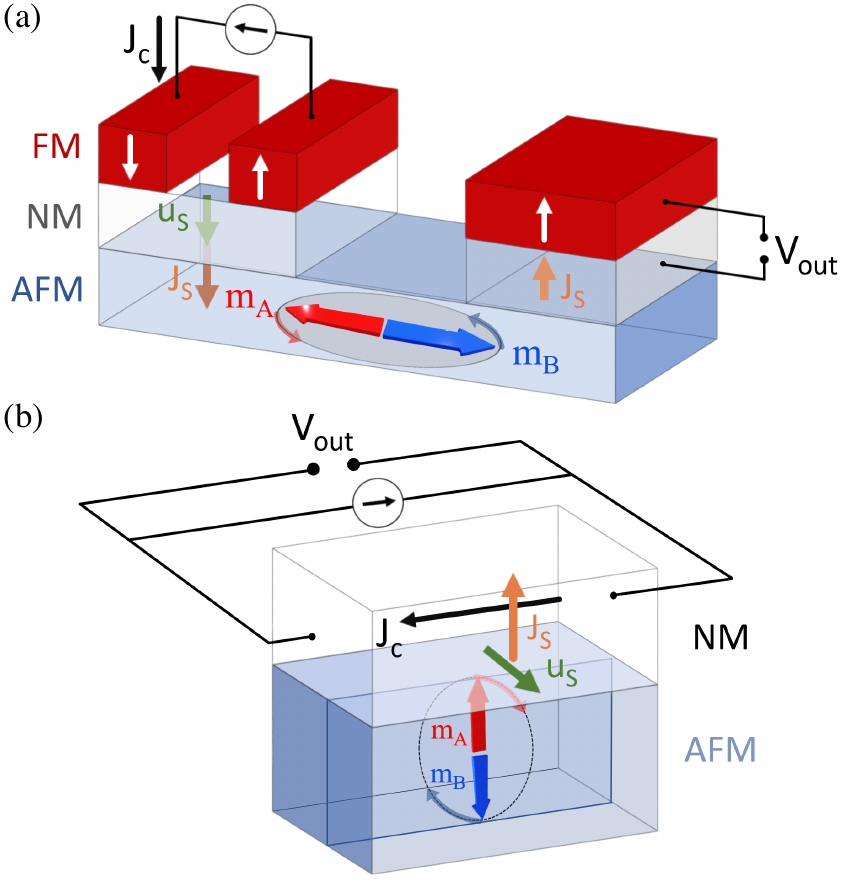}%
\caption{Device setup for electrical control and detection of terahertz oscillations in thin-film antiferromagnetic insulators. (a) In-plane spin valve geometry: lateral spin valve structure made of a perpendicular reference-layer magnetization on top of an in-plane-anisotropy antiferromagnet. (b) Perpendicular spin Hall geometry: spin Hall structure with in-plane electric current transverse to in-plane hard axis on top of a perpendicular-anisotropy antiferromagnet. Input electric current $J_\text{c}$ is converted into a pure spin current $J_\text{s}$ polarized $\vb{u}_\text{s}$ along the hard-anisotropy axis to cause precession of the sublattice moments $\vb{m}_A$ and $\vb{m}_B$, which pumps spin current back to generate an oscillating voltage signal at the output $V_\text{out}$.  FM: ferromagnet, NM: normal metal, AFM: antiferromagnet. }%
\label{fig:geo}
\end{figure}

Pure spin current through an insulating antiferromagnetic layer can be generated from electric current flowing in an overlaid lateral spin valve or spin Hall structure. In the lateral spin valve structure shown in Fig.~\ref{fig:geo}a, electric current across the ferromagnetic reference layers injects spin accumulation in the normal metal and spin current in the antiferromagnet, polarized parallel to the reference-layer magnetization~\cite{jedema2001electrical}, which can be oriented along the desired direction by a magnetic field. In the spin Hall structure shown in Fig.~\ref{fig:geo}b, electric current in the heavy metal with strong spin-orbit coupling produces a perpendicular spin current polarized transverse to electric current in the film plane~\cite{sinova2015}. The spin Hall effect can be efficient in converting electric current to spin, but the spin polarization is constrained within the plane. However, a spin-source material with reduced crystalline symmetry~\cite{macneill2017control} or planar Hall effect in a ferromagnet lifts this restriction~\cite{safranski2020planar}, which allows for controllable spin-polarization direction.

For the choice of spin polarization along the hard axis of the antiferromagnetic layer, the viable geometries are the in-plane-anisotropy antiferromagnet with the lateral spin valve structure made of a perpendicular reference-layer magnetization, and the perpendicular-anisotropy antiferromagnet with the spin Hall structure in which the in-plane electric current is transverse to the in-plane hard axis. For both the geometries, the input is a constant-current source, the back diffusion (``backflow") of injected spins~\cite{jiao2013spin} is considered small, and the output is detected under open-circuit condition.

To excite and control N\'eel-order precession, the spin current must exert adequate antidamping-like torque. The threshold spin-current density that initiates precession is obtained from the condition $\Gamma_\text{th}=1$ \eqref{eq:Gammam} as $J_\text{s,th} = \mathcal{K}_\text{e}d_\text{a}/2$. The corresponding electric-current density depends on the specific geometry. For the in-plane spin valve geometry, assuming low spin-memory loss in the normal metal, the conversion from electric current to spin is determined by the conductance of majority and minority-spin electrons $g_\text{M}$ and $g_\text{m}$, respectively, and the spin-mixing conductance at the interface of normal metal and antiferromagnet $g_\text{r}$. The threshold electric-current density is~\cite{skarsvaag2015spin}
\begin{equation}
J_\text{c,th}^{\text{ip-valv}} = \frac{2e}{\hbar} \frac{(g_\text{r}+g_\text{M}+g_\text{m})(g_\text{M}+g_\text{m})}{g_\text{r}(g_\text{M}-g_\text{m})} J_\text{s,th}.
\end{equation}
For the perpendicular spin Hall geometry, the conversion from electric current to spin is determined by the spin Hall angle $\Theta_\text{s}$, the layer thickness $d_\text{n}$, the spin diffusion length $\lambda$ and the conductivity $\sigma$ of the heavy metal, and the spin-mixing conductance at the interface of heavy metal and antiferromagnet $g_\text{r}$. The threshold electric-current density is~\cite{cheng2016dynamic}
\begin{equation}
J_\text{c,th}^{\text{pe-Hall}} = \frac{e}{\hbar} \frac{\sigma}{\lambda g_\text{r}} \frac{\coth{(d_\text{n}/2\lambda)}}{\Theta_\text{s}} J_\text{s,th}.
\end{equation}
If the effective damping $\beta \ll 1$, the minimum current for sustaining precession is about $1/\beta$ times lower as seen from Fig.~\ref{fig:conplot} and Eq.~\eqref{eq:Gammam}. Lowering the current drive after initiation mitigates excessive Joule heating in the metal, and allows for tunability of oscillation to sub-terahertz frequencies detectable by microelectronic circuits.

Precessing N\'eel order can reciprocally pump time-varying spin current $\propto \vb{n}\times\vb{n}'$ back into the adjacent metal, and experience a damping-like backaction~\cite{cheng2014spin}. This virtually enhances the Gilbert damping expressed as $\alpha = \alpha_0 + \alpha_\text{s}$, where $\alpha_0$ is the intrinsic damping constant and the enhancement~\cite{cheng2016dynamic}
\begin{equation}
\alpha_\text{s} = \frac{\hbar^2 \gamma g_\text{r}}{2 e^2 M_\text{s} d_\text{a}}.
\end{equation}
The pumped spin current is converted into voltage under open-circuit condition via spin filtering for the in-plane spin valve geometry and via inverse spin Hall effect for the perpendicular spin Hall geometry. The voltage signal generated at the output of each geometry is~\cite{skarsvaag2015spin, cheng2016terahertz}
\begin{gather}
V_\text{out}^{\text{ip-valv}}(t) = \frac{\hbar\gamma \sqrt{\mathcal{J}\mathcal{K}_\text{e}}}{2 e M_\text{s}} \frac{g_\text{r}(g_\text{M}-g_\text{m})\omega(t)}{(g_\text{r}+g_\text{M}+g_\text{m})(g_\text{M}+g_\text{m})},\\
V_\text{out}^{\text{pe-Hall}}(t) = \frac{\hbar\gamma \sqrt{\mathcal{J}\mathcal{K}_\text{e}}}{2 e M_\text{s}} \frac{\lambda g_\text{r}}{\sigma} \Theta_\text{s} \tanh(\frac{d_\text{n}}{2\lambda}) \omega(t),
\end{gather}
where $\omega$ is the dimensionless angular frequency (Sec.~\ref{Sec:angular}).

\begin{table}%[H] add [H] placement to break table across pages
 \caption{\label{tab:Physprop} Properties of bulk NiO and Cr\textsubscript{2}O\textsubscript{3} at 300 K}
 \begin{ruledtabular}
 \begin{tabular}{@{} l c r c r @{}}
Parameter & NiO & Ref. & Cr\textsubscript{2}O\textsubscript{3} & Ref. \\[1pt]
\hline\\[-7pt]
$\mathcal{J}$ (J/m\textsuperscript{3}) & $3.4\times 10^8$ & \onlinecite{sievers1963far},\onlinecite{khymyn2017antiferromagnetic} & $9.5\times10^7$ & \onlinecite{parthasarathy2019dynamics},\onlinecite{foner1963high} \\[1pt]
$\mathcal{K}_\text{e}$ (J/m\textsuperscript{3}) & $2.2\times 10^4$ & \onlinecite{sievers1963far},\onlinecite{khymyn2017antiferromagnetic}  & $3.2\times 10^3$ & \onlinecite{parthasarathy2019dynamics},\onlinecite{foner1963high} \\[1pt]
$\mathcal{K}_\text{h}$ (J/m\textsuperscript{3}) & $5.5\times 10^5$ & \onlinecite{sievers1963far},\onlinecite{khymyn2017antiferromagnetic} & $\approx 0$ & \onlinecite{parthasarathy2019dynamics},\onlinecite{foner1963high}\\[1pt]
$M_\text{s}$ (A/m) & $3.5\times 10^5$ & \onlinecite{hutchings1972measurement},\onlinecite{khymyn2017antiferromagnetic} & $1.9\times10^5$ & \onlinecite{parthasarathy2019dynamics},\onlinecite{foner1963high} \\[1pt]
$\alpha_0$ & $6\times 10^{-4}$ & \onlinecite{hutchings1972measurement},\onlinecite{khymyn2017antiferromagnetic} & $2\times 10^{-4}$ & \onlinecite{parthasarathy2019dynamics},\onlinecite{samuelsen1970inelastic}
\end{tabular}
\end{ruledtabular}
 \caption{\label{tab:geopara}  Characteristics of device geometry}
 \begin{ruledtabular}
 \begin{tabular}{@{} l c r @{}}
Parameter & Layer info & Ref. \\[1pt]
\hline\\[-7pt]
$g_\text{M} = 10^{10}$ (S/m\textsuperscript{2}) & NM/FM & \onlinecite{skarsvaag2015spin}   \\[1pt]
$g_\text{m} = 10^{9}$ (S/m\textsuperscript{2}) & NM/FM  & \onlinecite{skarsvaag2015spin}  \\[1pt]
$g_\text{r} = 10^{14}$ (S/m\textsuperscript{2}) & NM/AFM & \onlinecite{baldrati2018full}  \\[1pt]
$d_\text{a} = 5$ nm & AFM & \onlinecite{khymyn2017antiferromagnetic} \\[1pt]
$\sigma = 10^6$ (S/m) & NM (Pt) & \onlinecite{khymyn2017antiferromagnetic} \\[1pt]
$d_\text{n} = 20$ nm & NM (Pt) & \onlinecite{khymyn2017antiferromagnetic} \\[1pt]
$\lambda = 7$ nm & NM (Pt) & \onlinecite{khymyn2017antiferromagnetic} \\[1pt]
$\Theta_\text{s}=0.1$ & NM (Pt) & \onlinecite{khymyn2017antiferromagnetic} 
\end{tabular}
\end{ruledtabular}
\end{table}

Bipartite collinear antiferromagnets such as NiO and Cr\textsubscript{2}O\textsubscript{3} are insulators, whose bulk magnetic properties are well-studied (Table~\ref{tab:Physprop}). For thin films, density functional theory predicts that NiO(001) on SrTiO\textsubscript{3} substrate has an in-plane anisotropy~\cite{chen2018antidamping}, and experiments have shown that Cr\textsubscript{2}O\textsubscript{3}(0001) on Al\textsubscript{2}O\textsubscript{3} substrate has a perpendicular anisotropy~\cite{wu2011imaging}, but measurement of magnetic anisotropy of antiferromagnetic films is less explored~\cite{jungwirth2018multiple}. The magnetic anisotropy of thin films and multilayers can significantly differ from bulk magnetic materials due to epitaxial strain from substrate and reduced local symmetry of surface atoms~\cite{sander2004magnetic}. The magnetic anisotropy energy is augmented by magnetostrictive and surface anisotropies which can elicit an in-plane or a perpendicular spin orientation independent of the specific growth direction of the film~\cite{krishnan2016fundamentals}.  Nonetheless, we consider the bulk properties to  estimate specifications for the proposed setup.  

For characteristic values of geometry parameters (Table~\ref{tab:geopara}), the threshold current density $J_\text{c,th}^{\text{ip-valv}} \approx 2\times 10^7$ A/cm\textsuperscript{2} and $J_\text{c,th}^{\text{pe-Hall}} \approx 10^8$ A/cm\textsuperscript{2}; the effective damping $\beta^{\text{ip-valv}} = 0.18$ and $\beta^{\text{pe-Hall}} = 0.24$; the frequency scale is 1.4 THz for NiO and 0.5 THz for Cr\textsubscript{2}O\textsubscript{3}; the amplitude of alternating output signal $v_\text{out}^{\text{ip-valv}} = 70$ \textmu V and $v_\text{out}^{\text{pe-Hall}} = 0.4$ \textmu V at the threshold current~\footnote{The AC signal is obtained by calculating the coefficient of the sinusoidal term in Eq.~\ref{eq:omegaOD}}. The specifications are promising for practical demonstration of a terahertz signal emitter at nanoscale lengths.

Additionally, the setup can detect a terahertz radiation by phase-locking if the antiferromagnet's crystal symmetry allows for staggered-field N\'eel spin-orbit torque, which includes materials such as CuMnAs and Mn\textsubscript{2}Au~\cite{gomonay2018narrow}. 
By changing the spin polarization angle and the threshold spin-current density from $2 J_\text{s,th}/d_\text{a} \propto \alpha\sqrt{\mathcal{J}\mathcal{K}_\text{e}}$ to $\mathcal{K}_\text{e}+\mathcal{K}_\text{h}$~\eqref{eq:Gamma0vec}, hence the precession axis, the minimum angular frequency $\omega_0$ can be tuned from zero to $\sqrt{\kappa+1}$. This method proportionally shifts the center of the phase-locking interval. Frequencies below $\omega_0$ are cut off and this system acts as a high-pass filter operating in the terahertz range.

\section{Conclusion}

We study the equation of motion of the N\'eel order of an antiferromagnet under the action of spin torque for steady-state solution, which involves precession in a circular orbit about an axis oriented along a general direction. The required components of the spin-polarization vector in the orbital plane are explicit functions of the axis orientation. We analyze the dynamics by bisecting it into angular motion in the orbit and perturbation from the orbital plane. 

The angular motion reduces to equation of a damped-driven pendulum described by two parameters which represent the effective damping and drive. The parameter space is explored to identify regions of periodicity, hysteresis and infinite-period bifurcation. Solutions for oscillations in angular velocity are derived when the damping and drive are separately large, and a function for the fundamental frequency is found. The oscillations are sinusoidal for large drive and spike-like for large damping near threshold.

The perturbing elevation from the orbital plane reduces to equation of a damped harmonic oscillator with an effective spring constant. The stability condition introduces a minimum cutoff frequency for orbits, using which the threshold spin current is calculated. Orbital axes which deviate sufficiently from the hard axis have nonzero cutoff frequency, increasing monotonically with highest value for the easy-axis direction. The threshold is lowest along the hard axis, which varies non-monotonically with the angle from the hard axis for small damping. 

Finally, we propose device setups for electrical control and detection of terahertz oscillations in thin-film antiferromagnets. A certain practical challenge in achieving arbitrary precessional axis orientations is generating adequate spin-current density that overcomes the threshold, or reducing the magnetic anisotropy by strain-engineering to lower the current requirement.

% If in two-column frequency, this environment will change to single-column format so that long equations can be displayed. 
%\begin{widetext}
%$$\mbox{put long equation here}$$
%\end{widetext}

% Figures should be put into the text as floats. 
% Use the graphics or graphicx packages (distributed with LaTeX2e).
% Put the label that you will use with \ref{} command in the braces of the \label{} command.
%
% \begin{figure}
% \end{figure}

% \begin{table}
% \end{tabular}
% \end{table}

\begin{acknowledgments}
This work was supported partially by the Semiconductor Research Corporation and the National Science Foundation (NSF) through ECCS 1740136, and from the MRSEC Program of the NSF under Award Number DMR-1420073. ADK and EC acknowledge support by the Air Force Office of Scientific Research under Grant FA9550-19-1-0307.
\end{acknowledgments}

\bibliography{refs}

\end{document}